\documentstyle[aps,multicol,epsfig,eqsecnum]{revtex}

\newcommand\be{\begin{eqnarray}}
\newcommand\ee{\end{eqnarray}}
\newcommand\ba{\begin{array}}
\newcommand\ea{\end{array}}
\def\r{\rangle}
\def\l{\langle}
\def\T{{\rm Tr}}
\def\cH{{\cal H}}

\def\cE{{\cal E}}

\def\openone{{\it I}}

\begin{document}
\title{Realization of POVMs using measurement-assisted programmable quantum processors}
\author{M\'ario Ziman~$^{1,2}$ and Vladim\'\i r Bu\v zek~$^{1,2}$}
\address{
$^{1}$Research Center for Quantum Information, Slovak Academy of Sciences,
D\'ubravsk\'a cesta 9, 842 28 Bratislava, Slovakia \\
$^{2}$Faculty of Informatics, Masaryk University, Botanick\'a 68a,
602 00 Brno, Czech Republic}
\maketitle
\begin{abstract}
We study
possible realizations of generalized quantum measurements
on measurement-assisted programmable quantum processors.
We focus our attention on the realization of von Neumann measurements and
informationally complete POVMs. It is known that
two unitary transformations implementable by the same programmable processor
require mutually {\it orthogonal} states.
It turns out that the situation with von Neumann measurements is different.
Specifically, in order to realize two such measurements one does not
have to use orthogonal program states. On the other hand, the number of the implementable
von Neumann measurements is still limited. As an example of a programmable processor
we use the so-called quantum information distributor.
\end{abstract}
\begin{multicols}{2}
\section{Introduction}
General quantum measurements are formalized as {\it positive operator
valued measures} (POVM), i.e. sets of positive operators
$\{F_k\}$ summing up to identity, $\sum_k F_k=\openone$
(see, for instance, Refs.~\cite{Peres93,Preskill98,Nielsen2000,Busch}). From
the quantum theory it follows that each collection of such operators
corresponds to a specific quantum measurement. However, the theory does not tell
us anything about particular physical realization of a specific POVM.
The aim of this paper is to exploit the so-called
{\it measurement-assisted quantum processors} to perform
POVMs.

The {\it Stinespring-Kraus theorem} \cite{Stinespring1955}
relates quantum operations
({\it linear completely positive trace-preserving maps})
with unitary transformations.
In particular, any quantum operation $\cE$ realized on the system $A$
corresponds to a unitary transformation $U$ performed on a larger system
$A+B$, i.e.
\be
\cE[\varrho]=\T_B [G\varrho\otimes\xi G^\dagger]\; ,
\ee
where $\xi$ is a suitably chosen state of the system $B$
and $\T_B$ denotes a {\it partial trace} over the ancillary
system $B$. The assignment $\cE \mapsto (G,\xi)$ is one-to-many, because
the dilation of the Hilbert space of the system $A$ can be performed in many
different ways.
However, if we fix the transformation $G$,  states $\xi$ of the ancillary system
$B$ control and determine quantum operations that are going
to be performed on the system $A$. In this way one obtains a concept of a
{\it programmable quantum processor},
i.e. a fixed piece of hardware taking as an input a data register (system $A$)
and a program register (system $B$). Here the state of the program register
$\xi$  encodes the operation
$\varrho\to\varrho^\prime=\cE_\xi[\varrho]$ that is going to be performed
on the data register.

In a similar way, any quantum generalized measurement (POVM),
that is represented by a set of positive operators $\{F_j\}$,
can be understood as a {\it von Neumann measurement}
performed on the larger system \cite{Busch}.
von Neumann measurements
are those for which $F_j\equiv E_j$ are mutually orthogonal projectors,
i.e. $E_j E_k=\delta_{jk} E_k$. The {\it Neumark theorem} (see, e.g.
Ref.~\cite{Holevo1982})  states that
for each POVM $\{F_j\}$ there exists a von Neumann measurement
$\{E_j\}$ on a larger Hilbert space $\cH_{AB}$ and
$\T \varrho F_j=\T [(\varrho\otimes\xi) E_j]$ for all $\varrho$,
where $\xi$ is some state of the system $B$. Moreover, it is always
possible to choose a von Neumann measurement such that
$E_j= G^\dagger(I\otimes Q_j)G$ are $G$ is a unitary transformation  and
$Q_j$ are projectors defined on the system $B$.
Using the cyclic property of a trace operation, i.e.
$\T [(\varrho\otimes\xi)G^\dagger(I\otimes Q_j)G]=\T[G(\varrho\otimes\xi)G^\dagger(I\otimes Q_j)]$,
we see that the von Neumann measurement can be understood
as a unitary transformation $G$ followed by a von Neumann measurement
$M\leftrightarrow\{Q_j\}$ performed on the ancillary system only.

As a result we obtain the couple $(G,M)$ that determines a programmable quantum processor
assisted by a measurement of the program register, i.e.
{\it measurement-assisted programmable quantum processor}. Such device can be used to
perform both generalized measurements as well as quantum operations.

Programmable quantum processors (gate arrays of a finite extent) has been studied first
by Nielsen and Chuang \cite{Nielsen1997}. They have shown that no programmable quantum processor
can perform all unitary transformations of a data register.
To be specific, in order to encode $N$ unitaries into a program register one needs $N$ mutually orthogonal
program states. Consequently, the required program register
has to be described by an inseparable Hilbert space, because
the number of unitaries is uncountable. However,
if we work with a measurement-assisted programmable quantum processor,
then with a certain probability of success we can realize all unitary
transformations \cite{Vidal00,Hillery011,Hillery012,Hillery2002}.
The probability of success
can be increased arbitrarily close to unity utilizing
conditioned loops with a specific set of error correcting program
states \cite{Vidal00,Ziman03,Hillery04,Brazier2004}.

So far, the properties of quantum processors with respect to realization of quantum operations has been studied in several
papers \cite{Nielsen1997,Vidal00,Hillery011,Hillery012,Hillery2002,Vlasov2002}.
In the present paper we will exploit measurement-assisted quantum processors
to perform POVMs. The problem of the implementation
of von Neumann measurement
by using programmable ``quantum multimeters'' for discrimination of quantum
state has been introduced in Ref.~\cite{Dusek02} and subsequently studied in
Refs.~\cite{Fiurasek02,Fiurasek03,Bergou04}.
An analogous setting
of a unitary transformation followed by a measurement
has been used in Ref.~\cite{Paz03} to evaluate/measure the expectation
value of any operator. The quantum network based on a controlled-SWAP
gate can be used to estimate non-linear functionals of
quantum states \cite{Ekert02} without any recourse to quantum
tomography. Recently D`Ariano and co-wrokers \cite{DAriano03a,DAriano03b,DAriano04}
have studied how programmable quantum measurements can be efficiently realized with finite-dimensional ancillary systems.
In the present paper we will study how von Neumann measurements and informationally complete POVMs can be realized
via programmable quantum measurement devices. In particular, we will show that this goal can be achieved using
the so called quantum information distributor \cite{Braunstein2000,Rosko2003}.

\section{General consideration}
Let us start our investigation with an assumption that the program register
is always prepared in a pure state, i.e. $\xi=|\Xi\r\l\Xi|$. In this case the
action of the processor can be written in the following form
\be
G|\psi\r\otimes|\Xi\r=\sum_k A_k(\Xi)|\psi\r\otimes|k\r\; ,
\ee
where $|k\r$ is some basis in the Hilbert space of the program register and
$A_k(\Xi)=\l k|G|\Xi\r$. In particular, we can use the basis
in which the measurement $M$ is performed, i.e.
$Q_a=\sum_{k\in J_a}|k\r\l k|$, where $J_a$ is a subset of indices
$\{k\}$. Note that $J_a\cap J_{a^\prime}=\emptyset$, because
$\sum_a Q_a=\openone$.

Measuring the outcome $a$ the data evolve according to the following rule
({\it the projection postulate})
\be
\nonumber
\varrho\to\varrho^\prime_a &=&
\frac{1}{p_a}
\T_p [(\openone\otimes Q_a)G(\varrho\otimes|\Xi\r\l\Xi|)G^\dagger]\\
&=&
\frac{1}{p_a} \sum_{k\in J_a} A_k(\Xi)\varrho A_k^\dagger(\Xi)\; ,
\ee
with the probability
$p_a=\T [(\openone\otimes Q_a)G(\varrho\otimes|\Xi\r\l\Xi|)G^\dagger] =
\T[\varrho\sum_{k\in J_a}A_k^\dagger(\Xi) A_k(\Xi)]=\T[\varrho F_a]$.
Consequently for the elements of the POVM we obtain
\be
F_a=\sum_{k\in J_a} A_k^\dagger(\Xi) A_k(\Xi)\; .
\ee

If we consider a general program state with its spectral decomposition
in the form $\xi=\sum_n \pi_n |\Xi_n\r\l\Xi_n|$, then
the transformation reads
\be
\varrho\to\varrho_a^\prime= \frac{1}{p_a}
\sum_{n,k\in J_a} \pi_n A_{kn}\varrho A_{kn}^\dagger\; ,
\ee
with $A_{kn}=\l k|G|\Xi_n\r$ and
$p_a=\sum_{n,k\in J_a}\pi_n \T[\varrho A^\dagger_{kn}A_{kn}]$. Therefore
the operators
\be
F_a= \sum_{n,k\in J_a}\pi_n A^\dagger_{kn} A_{kn}
\ee
constitute the realized POVM.

Given a processor $G$ and some measurement
$M$ one can easily determine which POVM can be performed.
Note that the same POVM can be realized in many physically
different ways. Two generalized measurements $M_1,M_2$ are equivalent, if the
resulting functionals $f^{(x)}_k(\varrho)=\T\varrho F^{(x)}_k$ ($x=1,2$)
coincide for all $k$, i.e. they result in the same probability distributions.
For the purpose of the realization of POVMs, the state
transformation during the process is irrelevant. However, two equivalent
realizations of POVM can be distinguished by the induced state
transformations (for more on quantum measurement see Ref.~\cite{Busch}).

Let us consider, for instance, the trivial POVM, which consists of operators
$F_k= c_k\openone$ ($c_k\ge 0,\sum_k c_k=1$). In this case the observed
probability distribution is data-independent and some quantum operation is
realized. In all other cases, the state transformation
depends on the initial state of the data register, and is not linear
\cite{Ziman03}. In these cases the resulting distribution is nontrivial
and contains some information about the state $\varrho$. In the specific case
when the state $\varrho$ can be determined (reconstructed) perfectly, the measurement is
{\it informationally complete}. In this case we can perform the
{\it complete state reconstruction}. Any collection of $d^2$
linearly independent positive operators $F_k$ determine such informationally
complete POVM. In particular, they form an operator basis, i.e.
any state $\varrho$ can be written as a linear combination
$\varrho=\sum_j \varrho_j F_j$. Using this expression the probabilities
read
\be
\label{7}
p_j=\T [\varrho F_k]= \sum_k \varrho_k \T [F_j F_k] = \sum_k \varrho_k L_{jk}\; ,
\ee
where the coefficients $L_{jk}=\T [F_j F_k]$ define a matrix $L$.
In this setting the (inverse) problem
of the state reconstruction reduces to a solution of a system of linear
equations $p_j=\sum_k L_{jk}\varrho_k$, where $\varrho_k$ are unknown.
The solution exists only if the matrix $L$ is invertible and then
$\varrho_k = \sum_j L^{-1}_{kj} p_j$.

The purpose of any measurement is to
provide us information about the state of the physical system
based on the results of measurement.
The presented scheme of measurement-assisted quantum processor
represents quite general picture of the physical realization
of any POVM.

\section{Quantum information distributor}
In this section we will present a specific example of a quantum processor
the so-called {\it quantum information distributor} ({\tt QID}) \cite{Braunstein2000}. This device uses
as an input a
two-qubit program register and a single-qubit data register.
The processor consists of four {\tt CNOT} gates. Its name
reflects the property \cite{Braunstein2000} that in special
cases of  program states the
{\tt QID} acts as an optimal cloner and the optimal universal {\tt NOT} gate, i.e. it optimally
distributes quantum information according to a specific prescription. Moreover, it can
be used to perform an arbitrary qubit rotation with the probability
$p=1/4$ \cite{Hillery012}. The action of the {\tt QID} can be written in the form
\cite{Ziman03}
\be
\label{qid}
G_{\tt QID}|\psi\r\otimes|\Xi\r = \sum_k \sigma_k A(\Xi) \sigma_k|\psi\r\otimes|k\r\; ,
\ee
where $\sigma_k$ are sigma matrices, $A(\Xi)=\l k|G_{\tt QID}|\Xi\r$ and
$|k\r\in\{|0+\r,|1+\r,|0-\r,|1+\r\}$ is a two-qubit program-register basis
in which the measurement $M$ is performed
($|\pm\r=\frac{1}{\sqrt{2}}(|0\r\pm|1\r)$).

In what follows we shall extend the list of applications of
the {\tt QID} processor and show how to realize
a complete POVM, i.e. a complete state reconstruction.
For a general program state $|\Xi\r=\sum_k\alpha_k |\Xi_k\r$ with
$|\Xi_k\r=(\sigma_k\otimes\openone)|\Xi_0\r$
($|\Xi_0\r=\frac{1}{\sqrt{2}}(|00\r+|11\r)$) the POVM consists of
the following four operators
\be
F_k&=&\sigma_k F_{0+}\sigma_k=
\sigma_k A(\Xi)^\dagger A(\Xi)\sigma_k\; ,
\ee
with $F_{0+}=\frac{1}{4}\openone+\frac{1}{4}
[\alpha_0\vec{\alpha^*}+\alpha^*_0\vec{\alpha}+
i\vec{\alpha^*}\times\vec{\alpha}]
\cdot\vec{\sigma}$ and
$\vec{\alpha^*}=(\alpha_1^*,\alpha_2^*,\alpha_3^*)$,
$\vec{\alpha}=(\alpha_1,\alpha_2,\alpha_3)$.

Note that for the initial program state $|\Xi\r$ with
$\alpha_0=\cos\mu$,
$\vec{\alpha}=\frac{i\sin\mu}{\mu}\vec{\mu}$ ($\mu=||\vec{\mu}||$)
the probabilities $p_{0+}=\T F_{0+}\varrho=1/4$ are $\varrho$-independent,
and a unitary operation
$U_\mu=\exp(i\vec{\mu}\cdot\vec{\sigma})$ is realized \cite{Hillery012}.
The question of interest is whether
an informationally complete POVM can be encoded into
a program state. In fact,
the problem reduces to
the question of the linear independency of operators $F_k$ for some
$|\Xi\r$. Using the vector representation of operators,
$F_k=1/4(\openone+\vec{r}_k\cdot\vec{\sigma})$, one can
show that the operators $F_k$ are linearly independent only if
none of the coefficients of
$\vec{r}_{0+}=\alpha_0\vec{\alpha^*}+\alpha^*_0\vec{\alpha}+
i\vec{\alpha^*}\times\vec{\alpha}$ vanishes.

The elements of a POVM
can be represented in the Bloch-sphere picture. This is due to the fact that operators
$F_k=\frac{1}{2}\varrho_k$, and $\varrho_k$ represent quantum states.
Choosing the program state
\be
|\Xi_{POVM}\r=\frac{1}{\sqrt{2}}|\Xi_0\r+\frac{1}{\sqrt{6}}
(|\Xi_1\r+|\Xi_2\r+|\Xi_3\r)
\ee
we obtain the informationlly complete POVM with a very symmetric structure.
In particular, the operators $F_k$ are proportional to pure states
associated with vertexes  of a tetrahedron drawn inside
the Bloch sphere (see Fig.~1). These operators read
\be
F_{0+}&=&\frac{1}{4}(\openone+\frac{1}{\sqrt{3}}[\sigma_x+\sigma_y+\sigma_z])\; ;\\
F_{0-}&=&\frac{1}{4}(\openone+\frac{1}{\sqrt{3}}[-\sigma_x-\sigma_y+\sigma_z])\; ;\\
F_{1+}&=&\frac{1}{4}(\openone+\frac{1}{\sqrt{3}}[\sigma_x-\sigma_y-\sigma_z])\; ;\\
F_{1-}&=&\frac{1}{4}(\openone+\frac{1}{\sqrt{3}}[-\sigma_x-\sigma_y+\sigma_z])\; .
\ee
It is obvious that these operators are not mutually orthogonal,
but $\T F_j^\dagger F_k=\frac{1}{12}\delta_{jk}+\frac{1}{4}(1-\delta_{jk})$.
Using this identity one can easily compute the relation (\ref{7})
between the observed probability
distribution and the data state $\varrho$
\be
\varrho=\sum_k (-\frac{21}{5}p_k+\frac{9}{5}\sum_{j\ne k}p_j )|Q_k\r\l Q_k|\; ,
\ee
where we used the notation $F_k=\frac{1}{2}|Q_k\r\l Q_k|$.
The last equation completes the task of the state reconstruction task.
\begin{figure}
\label{fig1}
\begin{center}
\includegraphics[width=7cm]{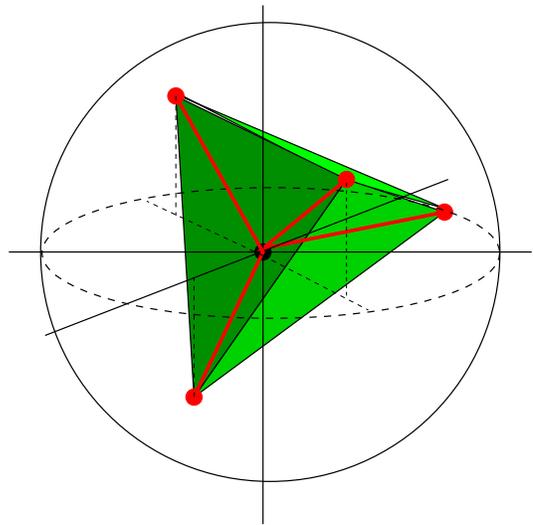}
\end{center}
\caption{
The Bloch sphere can be used to illustrate any POVM realized
on the QID processor. Each POVM is given by four operators that
determine four points in the Bloch sphere. Using this picture one can
see the structure and some properties of the realized POVM.
The depicted points forming a tetrahedron correspond to POVM elements
of the symmetric informationally complete POVM
associated with the program state
$|\Xi\r=\frac{1}{\sqrt{2}}|\Xi_0\r+\frac{1}{\sqrt{6}}
(|\Xi_1\r+|\Xi_2\r+\Xi_3\r)$.}
\end{figure}
Because of the identity ${\rm Tr} F_jF_k=const$ for
$j\ne k$ the realized POVM $\{F_k\}$ is of a special form.
It belongs to a family of the so-called symmetric informationally complete
measurements (SIC POVM) \cite{Caves}. These
measurements are of interest in several tasks
of quantum information processing and possess
many interesting properties. It is known (see. e.g. Ref.~\cite{Caves})
that for qubits there essentially exist only only two (up to unitaries)
such mesurements. Above we have shown how one of them can be
performed using the QID processor.

\section{von Neumann measurements}
An important class of  measurements is described
by the {\it projector valued measures} (PVM), which
under specific circumstances enable us to
distinguish between orthogonal states in a single shot, i.e.
no measurement statistics is required.
A set of operators $\{E_k\}$ form a PVM,
if $E_j=E_j^\dagger$ and $E_jE_k=E_j\delta_{jk}$, i.e. it contains mutually
orthogonal projectors. The total number of (nonzero) operators
$\{E_k\}$ cannot be larger than the dimension of the Hilbert space $d$.

Usually the {\it von Neumann measurements} are understood as those
that are compatible with the {\it projection postulate}, i.e. the result
$j$ associated with the operator $E_j=|e_j\r\l e_j|$ induces
the state transformation
\be
\varrho\to\varrho^\prime_j=\frac{E_j\varrho E_j}{\T\varrho E_j}=
\frac{|e_j\r\l e_j|\varrho|e_j\r\l e_j|}{\l e_j|\varrho |e_j\r}= |e_j\r\l e_j|=E_j\; .
\ee
That is, the state after the measurement is described by
the corresponding projector $E_j$.

However, each PVM
can be realized in many different ways and a particular von Neumann measurement
is only a specific case.
In our settings the realized POVM $\{F_k\}$ is related
to the state transformation via the identity
$F_k=A_k^\dagger A_k$, where
$\varrho\to\varrho^\prime_k=A_k\varrho A^\dagger_k$.
The set of operators $A_k=U_k E_k$,
with $E_k$ projectors and $U_k$ unitary transformations,
define the same PVM given by $\{E_k\}$. In particular,
$A_k^\dagger A_k= E_k U_k^\dagger U_k E_k = E_k E_k = E_k$, but
the state transformation results in
\be
\varrho\to\varrho^\prime_k=U_k E_k U_k^\dagger \ne E_k\; .
\ee
Thus the final state is described by a projector, but not in accordance
with the projection postulate. We refer to the
PVMs that are compatible with the projection postulate as the {\it von Neumann
measurements}. Moreover, for a simplicity we shall assume that
the projectors are always one-dimensional, i.e. the PVM is associated
with non-degenerate hermitian operators.

The action of the processor $G$ implementing two von Neumann measurements
$\{E_j\}$ and $\{G_j\}$
can be written as
\be
G|\psi\r\otimes|\Xi_E\r&=&\sum_j E_j|\psi\r\otimes|j\r\; ; \\
G|\psi\r\otimes|\Xi_G\r&=&\sum_j G_j|\psi\r\otimes|j\r\; .
\ee
It is well known \cite{Hillery2002} that when two sets of Kraus operators
are realizable by the same processor $G$, then
the following necessary relation holds
$\sum_j E_j G_j = \l\Xi_E|\Xi_G\r \openone$.
Using this relation for the projections $E_j=|e_j\r\l e_j|$,
$G_j=|g_j\r\l g_j|$ we obtain the identity
\be
\label{dd}
\sum_j u_{jj} |e_j\r\l g_j| =k \openone\; ,
\ee
where $u_{jj}=\l g_j|e_j\r$. For general measurements,
the operator on the left-hand side of the previous equation contains
off-diagonal elements. In this case the corresponding
program states must be orthogonal, i.e. $k=0$.
This result is similar to the one
obtained by Nielsen and Chuang \cite{Nielsen1997} who have studied
the possibility of the realization of unitary transformations via programmable
gate arrays.
They have shown that in order to perform (with certainty) two unitary transformations on a fixed
quantum processor one needs two orthogonal program states.
However, in our case we still cannot be sure that measurement-assisted
processor realizing two von Neumann measurements exists. Moreover, there are possibilities, when the condition holds also
for non-orthogonal program states (see the case study below).

In order to realize a projective measurement on a $d$-dimensional data register
the program space must be at least $d$ dimensional.
Let us start with the assumption that the Hilbert space of the program register is
$d$ dimensional. In this case the program states have to be
orthogonal (this is due to the fact the  expression (\ref{dd}) contain
off-diagonal elements). Let us consider $d$ different (non-degenerate)
von Neumann measurements
$M_\alpha$ that are determined by a set of operators
$E_k^\alpha=|\alpha_k\r\l\alpha_k|$
($E_k^\alpha E_j^\alpha=\delta_{kj} E_k^\alpha$ and
$\sum_k E_k^\alpha=\openone$ for all $\alpha$). Let $|\alpha\r$ denote
the associated program states and $\l\alpha|\beta\r=\delta_{\alpha\beta}$.
It is easy to see that for general measurements the resulting operator
\be
\label{ortpro}
G=\sum_{k,\alpha} E^\alpha_k\otimes|k\r\l\alpha|
\ee
is not unitary. In particular, $G^\dagger G =
\sum E_k^\beta E_k^\alpha \otimes |\beta\r\l\alpha| \ne \openone$.
The equality would require that the identity
$\sum_k E_k^\alpha E_k^\beta = \delta_{\alpha\beta}\openone_d $ holds.
Therefore, we conclude that neither orthogonal states
do guarantee the existence of a programmable processor that performs
desired set of measurements.  This result makes the case of
programming the unitaries and von Neumann measurements different.

\begin{table}
\label{table1}
\be
\nonumber
\ba{|l||c|c|c|c|}
\hline
{\rm measurement} & M_1 & M_2 & \dots & M_d \\ \hline\hline
{\rm result}\ 1 & |\alpha_1\r & |\beta_1\r & \dots &|\omega_1\r \\ \hline
{\rm result}\ 2 & |\alpha_2\r & |\beta_2\r & \dots &|\omega_2\r \\ \hline
\vdots & \vdots & \vdots & \vdots & \vdots \\ \hline
{\rm result}\ d & |\alpha_d\r & |\beta_d\r & \dots & |\omega_d\r \\ \hline
\ea
\ee
\caption{The measurements $M_1,M_2,\dots,M_d$
are realizable by a $d$ dimensional program register
only if all vectors in the rows are mutually orthogonal.
Moreover, no two columns can be related by a permutation.
The orthogonality of the vectors in columns is ensured by the
fact that they form a PVM. It turns out that the number
of realizable measurements equals at most to $d-2$, i.e. neither for
qutrit one can encode more than a single von Neumann measurement.
Moreover, the measurements that can be performed are not arbitrary. }
\end{table}

For instance, let us
consider a two-dimensional program register and let us denote
$E_{0,1}^0=E_{0,1}$ and $E_{0,1}^1=G_{0,1}$. Then the above condition
reads $E_0G_0=E_1G_1=0$. Using the definition $E_k=|e_k\r\l e_k|$
and $G_k=|g_k\r\l g_k|$ we obtain the orthogonality conditions
$\l e_0|g_0\r=\l e_1|g_1\r=0$. Consequently, because in the
two-dimensional case the orthogonal
state is unique, we obtain $|g_0\r = |f_1\r$
and $|g_1\r=|e_0\r$, i.e. the measurements are the same. For larger
$d$ the situation is different. The realizable measurements
must possess the derived property which can be summarized with the help
of Tab.~I.

In order to realize more von Neumann measurements on a qudit one
has to work with a larger-dimensional program space, i.e.
$\dim\cH_p=d_p>d$. In general, in this case we work with $d_p$
outcomes and $d_p$ projective operators $Q_k$ that define the
realized measurement of the program register. However, each PVM
consists of maximally $d$ projectors. Therefore, $d_p-d$ of the
induced operators $E_k$ should represent the {\it zero operator}.
It means that when we are realizing the von Neumann measurement
such that some of the outcomes do not occur, i.e. probability of
them vanishes for all data states. However, there is one more
option that the set of operators $\{E_k\}$ ($k=1,\dots, d_p$)
contains exactly only $d$ different operators (projectors). This
means that more results can specify the same projection and define
a single result of the realized von Neumann measurement.

The idea of additional, the so called, ``zero'' operators can be used to formulate
a general statement about the implementation of any collection of
arbitrary von Neumann measurements.
Let us consider $N$ von Neumann measurements $M_\alpha$ given by
non-zero operators $\{E_k^{\alpha}\}$ (number of $k$ equals to $d$).
We can define the sets of $d_p$ operators $\{\tilde{E}_k^\alpha\}$
by adding to these sets zero operators so that the condition $\sum_k \tilde{E}_k^\alpha
\tilde{E}_k^\beta=\delta_{\alpha\beta}\openone$ holds.
Using this approach we find out that any collection of $N$
von Neumann measurements can be realized on a single quantum processor
given by Eq.(\ref{ortpro}) with (maximally) $Nd$ dimensional program space.

\subsection{{\bf Case study:} {\it Projective measurements on a qubit.}}
Let us consider two von Neumann measurements $M=\{E_0,E_1\}$ and $N=\{G_0,G_1\}$ on a qubit.
Further, let us assume a three-dimensional program space and define measurements
$M_1=\{E_0,E_1,0\}$ and $M_2=\{0,G_1,G_2\}$, respectively. It is easy to see that neither
of these two sets of operators do satisfy the condition
$0=\sum_k E_kG_k= 0E_0+E_1 G_1+0G_2=E_1 G_1$. The equality holds only
if $E_1 G_1=0$, i.e. $E_1=|\psi\r\l\psi|$ and
$G_2=|\psi_\perp\r\l\psi_\perp|$, but this implies that both measurements
are the same. Consequently, the dimension of the program space
has to increase by one. Then we have $M_1=\{E_0,E_1,0,0\}$,
$M_2=\{0,0,G_0,G_1\}$ and the condition holds for all possible measurements
$M_1,M_2$. It follows that the implementation of $N$ von Neumann measurements
on a qubit requires $N$-qubit program space.

The program space of the {\tt QID} processor given by Eq.(\ref{qid})
consists of two qubits. Using the conclusion of the previous paragraph
it follows that two von Neumann measurements
could be performed with the help of this processor. It is easy to see
that the operators $A_k=\sigma_kA(\Xi)\sigma_k$ with
$A(\Xi)=\frac{1}{2}\sum_j \alpha_j\sigma_j$ are not
projectors. Moreover, they do not vanish only for certain $j$.
Consequently, the projective measurement cannot be realized
in the same way as described above.
However, the {\tt QID}-processor can still be exploited to perform
a von Neumann measurement.

Using the program state $|\Xi\r=\frac{1}{\sqrt{2}}(|\Xi_0\r+|\Xi_1\r)$ the
operator $A=\frac{1}{2\sqrt{2}}[\openone+\sigma_x]$ (i.e.,
$F_0=A^\dagger A=\frac{1}{2}P_+$, where $P_+=\frac{1}{2}[\openone+\sigma_x]$)
is a projection onto the vector $|+\r=\frac{1}{\sqrt{2}}{|0\r+|1\r}$.
It is obvious that $F_1=\sigma_x F_0\sigma_x=F_0$
and $F_2=F_3=\frac{1}{2}P_-$, where $P_-=\frac{1}{2}[I-\sigma_x]$. It turns
out that we have realized PVM described by $P_\pm$, i.e. the eigenvectors
of the $\sigma_x$ measurement. The state transformation reads
$\varrho\to \varrho_k^\prime=P_\pm$ (if $p_k\ne 0$), respectively.
It follows that the realization of
the measurement of $\sigma_x$ is in accordance with the projection
postulate. In the same way we can realize $\sigma_y$ and
$\sigma_z$ measurement (in these cases different results must be paired). Basically, this
corresponds to a choice of different two-valued measurements, but
in reality we perform only a single four-valued measurement.
As a result we find that on {\tt QID} we can realize three different
von Neumann measurements. Note that we have used only two qubits as the program
register. Moreover, the associated program states
$|\Xi_{\sigma_j}\r=\frac{1}{\sqrt{2}}[|\Xi_0\r+|\Xi_j\r]$
are not mutually orthogonal, but
$\l\Xi_{\sigma_j}|\Xi_{\sigma_k}\r=\frac{1}{2}$ (for $j\ne k$)
and  Eq. (\ref{dd}) holds.
Namely, for the measurements of
$\sigma_x\leftrightarrow\{P_\pm\}$ and
$\sigma_z\leftrightarrow\{P_0=|0\r\l 0|,P_1=|1\r\l 1|\}$
the condition (\ref{dd}) reads
$\frac{1}{2}[P_+P_0+P_+P_1+P_-P_1+P_-P_0]=\frac{1}{2}\openone$.

Till now we have always assumed that program states that encode two
von Neumann measurements have to be orthogonal. The last
paragraph describes an counterexample. As we already mentioned in some specific cases the
condition of the orthogonality can be relaxed.

\subsection{{\bf Note.} {\it Projection valued measures}}
If we relax the compatibility with the projection postulate more PVMs
can be realized on a single processor. Let us consider that the dimension of
the program space equals $d$ and $|\alpha\r$ is the state that
encodes the PVM given by a set $\{E_k^\alpha\}$. The action of $G$ can be written
as
\be
G|\psi\r\otimes|\alpha\r=\sum_k U_k^\alpha E_k^\alpha |\psi\r\otimes|k\r
\ee
and the condition
$\sum_k E_k^\alpha U_k^{\alpha\dagger} U_k^\beta E_k^\beta = \delta_{\alpha\beta}\openone$ must hold.
Let us consider two PVMs on a qubit
$\{E_0=|0\r\l 0|,E_1=|1\r\l 1|\}$
and $\{G_0=|\phi\r\l\phi|,G_1=|\phi_\perp\r\l\phi_\perp|\}$.
Define a unitary map $U$ such that $|\phi\r\to|1\r$
and $|\phi_\perp\r\to|0\r$. Using this map we can define
a processor by following equations
\be
G|\psi\r\otimes|\Xi_E\r &=& E_0|\psi\r\otimes|0\r+E_1|\psi\r\otimes|1\r\; ;
\nonumber\\
G|\psi\r\otimes|\Xi_G\r &=& \tilde{G}_0|\psi\r\otimes|0\r+\tilde{G}_1|\psi\r\otimes|1\r\; ,
\ee
where $\tilde{G}_0= U G_0=|1\r\l\phi|$,
$\tilde{G_1}=UG_1=|0\r\l\phi_\perp|$ and $\l\Xi_E|\Xi_G\r=0$.
Direct calculation shows that
$E_0\tilde{G}_0+E_1\tilde{G}_1=|0\r\l 0|1\r\l\phi|+|1\r\l 1|0\r\l\phi_\perp|=0$, i.e. $G$ is unitary. From here it follows that if
one does not require the validity of the projection postulate, then any two
PVMs can be performed on a processor with two-dimensional program space.

This result holds in general. Let us consider a set of $d$ PVMs
$\{E_k^\alpha\}$ on a qudit. There always exist unitary
transformations $U^\alpha$ such that operators $\tilde{E}_k^\alpha
= U^\alpha E_k^\alpha$ satisfy the condition $\sum_k
\tilde{E}_k^{\alpha\dagger} \tilde{E}_k^\beta=
\delta_{\alpha\beta}\openone$. Without the loss of generality we
can consider that measurement $M_0$ is given by projectors $|0\r\l
0|,\dots,|d-1\r\l d-1|$ and $M_\alpha$ by
$|\phi_0^{\alpha}\r\l\phi_0^\alpha|,\dots,
|\phi_d^{\alpha}\r\l\phi_d^\alpha|$ (see Tab.~II.).

\section{Conclusion}
In this paper we have studied how POVMs can be physically realized
using the so-called measurement-assisted quantum processors.
In particular, we have analyzed how to perform complete state reconstruction
and von Neumann measurements. As a result we have found
that an arbitrary collection of von Neumann measurements cannot be realized
on a single programmable quantum processor.
We have shown how to use the {\tt QID} processor
to perform the state reconstruction.

The number of implementable von Neumann measurements
is limited by the dimensionality of the program register.
Our main result is that with a program register composed of $N$ qudits one can surely
define a processor which performs arbitrary $N$ von Neumann measurements. In fact,
in general one can do much better. We have shown that
the usage of non-orthogonal program states can be helpful.
In particular, the {\tt QID} processor can be
exploited to perform three von Neumann measurements by using only
two qubits as a program register and nonorthogonal states.
Using only $d$ dimensional program space
one can realize maximally $N=d-2$ von Neumann measurements on a qudit
(for a qubit we have $N=1$).

Relaxing the condition on compatibility  with the projection postulate
the processor allows us to realize any collection of $N$ PVMs
just by using $d_p=N$ dimensional program space. An open question is whether
we can perform more PVMs, or not. The two tasks can be performed by programmable processors:
the realization
of von Neumann measurements and the application unitary transformations on the data register, are different.
According to Nielsen and Chuang \cite{Nielsen1997}, any collection
of $N$ unitary transformations requires $N$ dimensional program space.
For $N$ von Neumann measurements the
upper bound reads $d_p=Nd$ and any improvement strongly depends on
the specific set of these measurements. The characterization of these
classes of measurements is an interesting topic that will be studied elsewhere.

\acknowledgements
This work was
supported in part by the European Union  projects QUPRODIS, QGATES
and CONQUEST. We thank Peter \v Stelmachovi\v c for valuable discussions. We would also
like to thank Mari\'a n Ro\v sko for reading the manuscript and having no comments.

\begin{table}
\label{table1}
\be
\nonumber
\ba{|l||c|c|c|c|}
\hline
{\rm measurement} & M_1 & M_2 & \dots & M_d \\ \hline\hline
{\rm result}\ 1 & |\alpha_1\r & |\beta_1\r & \dots &|\omega_1\r \\ \hline
{\rm result}\ 2 & |\alpha_2\r & |\beta_2\r & \dots &|\omega_2\r \\ \hline
\vdots & \vdots & \vdots & \vdots & \vdots \\ \hline
{\rm result}\ d & |\alpha_d\r & |\beta_d\r & \dots & |\omega_d\r \\ \hline
\ea
\ee
\caption{The measurements $M_1,M_2,\dots,M_d$
are realizable by a $d$ dimensional program register
only if all vectors in the rows are mutually orthogonal.
Moreover, no two columns can be related by a permutation.
The orthogonality of the vectors in columns is ensured by the
fact that they form a PVM. It turns out that the number
of realizable measurements equals at most to $d-2$, i.e. neither for
qutrit one can encode more than a single von Neumann measurement.
Moreover, the measurements that can be performed are not arbitrary. }
\end{table}


\end{multicols}

\begin{thebibliography}{10}
\bibitem{Peres93}
A.Peres, {\it Quantum Theory: Concepts and Methods} (Kluwer,
Dordrecht, 1993).

\bibitem{Preskill98}
J.Preskill, {\it Quantum theory of Information and Computation},
{\tt www.theory.caltech.edu/people/preskill}

\bibitem{Nielsen2000}
M.A.Nielsen and I.L.Chuang {\it Quantum Computation and Quantum
information} (Cambridge University Press, Cambridge, 2000).

\bibitem{Busch}
P. Busch, P. Lahti, and P. Mittalstead, {\it Quantum Theory of
Measurement} (Lecture Notes in Physics m2, Springer Verlag, 1996).

\bibitem{Stinespring1955}
D.E. Evans and J.T. Lewis, {\it Dilations of Irreversible
Evolutions in Algebraic Quantum Theory}, Communications of Dublin
Institute of Advanced Studies, Series A (Theoretical Physics), No.
24, Dublin, DIAS (1977).

\bibitem{Holevo1982}
A.S. Holevo, {\it Probabilistic and Statistical Aspects of Quantum
Theory}
(North-Holland, Amsterdam, 1982).

\bibitem{Nielsen1997}
M. A. Nielsen and I. L. Chuang, {\it Programmable quantum gate
arrays}, Phys. Rev. Lett. {\bf 79}, 321 (1997).

\bibitem{Vidal00}
G. Vidal, L. Masanes,  and J.I. Cirac, {\it Storing quantum
dynamics in quantum states: A stochastic programmable gate},
Phys.Rev.Lett. {\bf 88}, 047905 (2002).

\bibitem{Hillery011}
M. Hillery, V. Bu\v zek, and M. Ziman, {\it Programmable quantum
gate arrays}, {Fortschritte der Physik} {\bf 49}, 987 (2001).

\bibitem{Hillery012}
M. Hillery, V. Bu\v zek, and M. Ziman, {\it Probabilistic
implementation of quantum processors}, Phys.Rev A {\bf 65}, 022301
(2002)

\bibitem{Hillery2002}
M. Hillery, M. Ziman, and V. Bu\v zek, {\it Implementation of
quantum maps by programmable quantum processors}, { Phys.Rev A}
{\bf 66}, 042302 (2002).

\bibitem{Ziman03}
M. Ziman and V. Bu\v zek, {\it Realization of unitary maps via
programmable quantum processors}, { Int. Journal of Quantum
Inf.} {\bf 1}, 527 (2003).

\bibitem{Hillery04}
M. Hillery, M. Ziman, and V. Bu\v zek, {\it Improving
performance of probabilistic programmable quantum processors},
Phys.Rev.A {\bf 69} (2004).

\bibitem{Brazier2004}
A. Brazier, V. Bu\v{z}ek, and P.L. Knight, {\it Probabilistic
programmable quantum processors with multiple copies of program
state}, submitted to  Phys.Rev.A.

\bibitem{Vlasov2002}
A. Yu. Vlasov, {\it Aleph-QP:Universal Hybrid quantum
processors}, {\tt quant-ph/0205074}.

\bibitem{Dusek02}
M. Du\v sek and V. Bu\v zek, {\it Quantum-controlled measurement
device for quantum-state discrimination}, Phys. Rev. A {\bf 66},
022112 (2002).

\bibitem{Fiurasek02}
J. Fiur\'a\v sek, M. Du\v sek, and R. Filip, {\it Universal
measurement apparatus controlled by quantum software},
Phys. Rev. Lett. {\bf 89}, 190401 (2002).

\bibitem{Fiurasek03}
J. Fiur\'a\v sek and M. Du\v sek, {\it Probabilistic quantum
multimeters}, Phys. Rev. A {\bf 69}, 032302 (2004).

\bibitem{Bergou04}
J.A. Bergou, M. Hillery, and V. Bu\v zek,
{\it Programmable quantum state discriminator with simple programs},
unpublished.

\bibitem{Paz03}
J.P. Paz and A.  Roncaglia, {\it A quantum gate array can be
programmed to evaluate the expectation value of any operator},
Phys. Rev. A {\bf 68}, 052316 (2003).


\bibitem{Ekert02}
A.K. Ekert, C.M. Alves, D.K.L. Oi, M. Horodecki, P. Horodecki, and
L.C. Kwek, {\it Direct estimation of linear and non-linear
functionals of a quantum state}, Phys. Rev. Lett. {\bf 88}, 217901 (2002).

\bibitem{DAriano03a}
G.M. D'Ariano, P. Perinotti, and M.F. Sacchi,
{\it Quantum universal detectors},
Europhys. Lett. {\bf 65}, 165 (2004).

\bibitem{DAriano03b}
G.M. D'Ariano, P. Perinotti, and M.F. Sacchi,
{\it Optimization of quantum universal detectors},
in "Proceedings of the 8th Int. Conf. on Squeezed States and Uncertainty Relations", ed. by H. Moya-Cessa et al., (Rinton, Princeton, 2003) p. 86.

\bibitem{DAriano04}
G.M. D'Ariano and P. Perinotti,
{\it Efficient universal programmable quantum measurements},
{\tt quant-ph/0410169}.

\bibitem{Braunstein2000}
S. Braunstein, V. Bu\v{z}ek, and  M. Hillery, {\it Quantum
Information Distributor:Quantum network for symmetric and
anti-symmetric cloning in arbitrary dimension and continuous
limit}, { Phys. Rev. A} {\bf 63}, 052313 (2001).

\bibitem{Rosko2003}
M. Ro\v{s}ko, V. Bu\v{z}ek, P.R. Chouha, and M. Hillery,
{\it Generalized measurements via programmable quantum processor},
Phys. Rev. A {\bf 68}, 062302 (2003).

\bibitem{Caves}
J.M. Renes, R. Blume-Kohout, A. J. Scott , and C.M. Caves,
{\it Symmetric informationally complete quantum measurements},
J. Math. Phys. {\bf 45}, 2171 (2004).

\end{thebibliography}
\end{document}